\documentclass{article}
\usepackage[T1]{fontenc}		
\usepackage[utf8]{inputenc}		

\usepackage{lmodern} 

\usepackage{authblk}
\usepackage{amssymb}
\usepackage{verbatim}
\usepackage{bm}
\usepackage{cite}
\usepackage{graphicx}
\usepackage{color}
\usepackage[colorlinks=true, allcolors=blue]{hyperref}
\usepackage{physics}
\usepackage{indentfirst}
\usepackage{braket}
\usepackage{cancel}
\usepackage{soul}

\usepackage{lastpage}			
\usepackage{indentfirst}		
\usepackage{color}			
\usepackage{graphicx}			
\usepackage{microtype} 
\usepackage{hyperref} 
\usepackage{xurl}
\usepackage{pdfpages}
\usepackage{verbatim} 

\usepackage{amsthm} 
\usepackage{amsmath} 
\usepackage{amsfonts} 
\usepackage{mathrsfs} 
\usepackage{mathtools} 
\usepackage{exscale} 
\usepackage{relsize} 
\DeclarePairedDelimiterX{\inp}[2]{\langle}{\rangle}{#1, #2}
\usepackage{dsfont}
\usepackage[normalem]{ulem}

{
\theoremstyle{definition}	
\newtheorem{definition}{Definition}[section]
\newtheorem{obs}{Remark}[section]
\newtheorem{example}{Example}[section]
}

\newtheorem{lemma}{Lemma}[section]
\newtheorem{theorem}{Theorem}[section]

\newcommand{\eft}{\mathcal{D} (\mathds{R}^n)}
\newcommand{\eftt}{\mathcal{D} (\mathds{R}^3)}
\newcommand{\eftsub}[1]{\mathcal{D} (#1)}
\newcommand{\eftomega}{\mathcal{D}_\omega (\mathds{R}^n)}
\newcommand{\eftdual}{\mathcal{D}' (\mathds{R}^n)}
\newcommand{\efttdual}{\mathcal{D}' (\mathds{R}^3)}
\newcommand{\eftdualsub}[1]{\mathcal{D}' (#1)}
\newcommand{\eftdualomega}{\mathcal{D}_\omega' (\mathds{R}^n)}

\providecommand{\sin}{} \renewcommand{\sin}{\hspace{2pt}\mathrm{sen}}

\newcommand{\norma}[1]{\left\vert #1 \right\vert}

\definecolor{Red}{rgb}{0.9,0,0}

\title{A comment on the classical electron self-energy}
\author[1]{H. R. de Assis\footnote{Corresponding author: heitor.ribeiro@estudante.ufjf.br}}
\author[2]{B. F. Rizzuti\footnote{Contact: brunorizzuti@ice.ufjf.br}}

\affil[1]{Departamento de Matemática, ICE, Universidade Federal de Juiz de Fora, MG, Brazil}

\affil[2]{Departamento de F\'isica, ICE, Universidade Federal de Juiz de Fora, MG, Brazil}

\date{}                     
\begin{document}

\maketitle

\begin{abstract}

This paper is devoted to the analysis of the divergence of the electron self-energy in classical electrodynamics. To do so, we appeal to the theory of distributions and a method for obtaining corresponding extensions. At first sight, electrostatics implies a divergence once we treat the electron as a charged point particle. However, our construction shows that its self-energy turns out to be an undetermined constant upon renormalization. Appealing to empirical results we may fix its value, demanding, for example, that all its mass comes from an electrostatic origin. 

\textbf{Keywords}: Theory of distributions. Extension of distributions. Electron self-energy.
\end{abstract}

\section{Introduction} \label{Sec1}

One of the most compatible matches between theory and experiment in physics can be seen in the field of Quantum Electrodynamics (QED), as the precision on the electron magnetic moment goes far from expected \cite{zee_quantum_2010}. Since the early days of QED computations, it became clear that one would have to deal with singularities, related, for example, with self-interactions. In turn, this has led the community to stare at fields not as maps, but as distributions. In the case of the electric field $\vec{E}(\vec{x},t)$ originated from a point particle, for instance, one would expect an ultraviolet divergence at origin, while $\int d^3\vec{x} dt \vec{E}(\vec{x},t) f(\vec{x}) = \vec{E}(f)$ is well behaved \cite{streater_pct_2016}. Here, $f(\cdot)$ is a smooth function of compact support. We are interested, in this manuscript, on the self-energy of the electron. While it has a fascinating history so depicted in \cite{Bovy:2006dr}, involving an entire war and a new generation of physicists developing regularization and renormalization techniques, the classical counterpart is often subdue, justifying our approach here.

Simply put, the self-energy of a charged particle, such as the electron, is the measure of the energy it posses when freed from any other interaction, be it with other particles or with given fields. One finds in the study of classical electrodynamics that, summed to the kinetic and potential energies a given collection of particles might have, a system composed of \textit{charged} particles has a quantity of energy related to the electromagnetic field it generates \cite{griffiths_electro_1999}.

The electromagnetic system we wish to examine could be seen, at first, as the simplest case: that of an electron, stationary, freed from any other interaction. Seen as a \textit{point particle} - that is, supposing it has no internal structure and being solely described by the position in which its whole charge is stored - which is the standard way one encounters at first \cite{griffiths_electro_1999,jackson_classical_1975}, the electric field and potential are given by
\begin{equation} \label{eq1.2}
    \textbf{E}(\textbf{r}) = \frac{1}{4 \pi \epsilon_0} \frac{e}{r^2} \hat{\textbf{r}}, \qquad V(\textbf{r}) = \frac{1}{4 \pi \epsilon_0} \frac{e}{r},
\end{equation}
where $e$ denotes the strength of its charge.

Meanwhile, the expression for the self-energy for a system with electric field $\textbf{E}$ and magnetic field $\textbf{B}$ is
\begin{equation} \label{eq1.6}
    E = \frac{\epsilon_0}{2} \int_{\mathds{R}^3} \left( \textbf{E}^2 + c^2 \textbf{B}^2 \right) d \tau,
\end{equation}
so that, using \eqref{eq1.2}, we obtain
\[ E_0 = \frac{\epsilon_0}{2} \int_{\mathds{R}^3} \left(\frac{1}{4 \pi \epsilon_0}\right)^2 \left( \frac{e}{r^2} \right)^2 d \tau. \]
We denote $E_0$ the self-energy of interest here, as we are allegedly neglecting the magnetic field due to our interest only in the static case.  
Using spherical coordinates,\footnote{Since different materials might use different notations concerning the polar coordinates $\theta$ and $\phi$, we make explicit that we are considering here $\begin{cases}
        x = r \sin{\theta} \cos{\phi}, \\
        y = r \sin{\theta} \sin{\phi}, \\
        z = r \cos{\theta},    
\end{cases}$
where $\theta \in [0,\pi]$, $\phi \in [0,2 \pi)$.}
\begin{equation}
    \begin{split}
        E_0 & = \frac{e^2}{(4 \pi \epsilon_0)^2} \left( \int_0^{2 \pi} d \phi \right) \left( \int_0^{\pi} \sin{(\theta)} d \theta \right) \left( \int_0^{+\infty} \frac{1}{r^2} d r \right) \\
        & = \frac{e^2}{8 \pi \epsilon_0} \left[ \frac{1}{r} \right]_{+\infty}^0 \\
        & = + \infty .
    \end{split}
\end{equation}

The conclusion we arrive then is that there is an infinite amount of energy stored in the field of a simple electron positioned at the origin of our system, if one considers it to be a stationary point particle. Needless to say, any satisfactory field theory, both classical or quantum, must resolve such type of divergences. Should we discard the assumption that the electron has no spatial extension? Both theoretical and experimental results seem to point in an opposite way \cite{hudson_improved_2011, acme_collaboration_improved_2018}, indicating that we should seek a improvement in the theory and in the conception of self-energy itself.

Therefore, here we present one of the available methods for the ``removal'' of such infinite quantities, a process known as \textit{renormalization} \cite{griffiths_introduction_2011}. The main idea of renormalization is that these infinities can be justified by attributing them to quantities which we cannot directly measure (something that can be seen as a parallel with the acceptance of complex numbers in the formalism of quantum mechanics). Take, for example, our case of the electron and its infinite self-energy. In calculating $E_0$, we are, simultaneously, calculating the mass $m_{elec}$ arising from the electric field of such particle, since Einstein's relativity theory affirms that mass and energy are but two manifestations of the same phenomenon. In light of this, we conclude that the divergence of $E_0$ implies that the electron possesses an infinite inertia. If, however, we assume there exists another contribution for the \textit{effective mass} (that is, the one we can actually measure), originated from some unknown effect other than electromagnetism, then we might conceive that this new contribution is negative enough to oppose the infinity appearing from $m_{elec}$. The hypothesis of another source contributing for the effective mass is not something hardly justified, since we know that neutral bodies also have mass and generate no electric or magnetic field. Thus, assuming this new contribution for the mass of the particle, independent of where it comes from, we can ``erase'' the infinite we have just found, obtaining the so called \textit{mass renormalization} of the electron. It might be difficult to accept a so far unknown (non-electromagnetic) interaction which is responsible for a strong renormalization of the mass, without leaving a fingerprint in the dynamical particle sector. Our effective result shows that a finite mass/energy is obtained, though, even without invoking any other renormalization procedure.

Such a primitive method for renormalization gave rise to new and more advanced approaches which came to be used in the renormalization of other infinite quantities, specially in Quantum Field Theory (QFT), which advanced quite a lot in the decades following the emergence of quantum mechanics and which presented similar problems with divergent integrals in its equations \cite{ryder_quantum_1996}. For such a new theory, the methods had to be refined in its mathematical formulation and, meanwhile, the development of approaches such as \textit{constructive quantum field theory} \cite{streater_pct_2016} and \textit{causal perturbation theory} (CPT) \cite{bogoliubov_introduction_1980,epstein_glaser_1973} made clear that distribution theory was a crucial subject for the understanding of such new ideas. It was also through this theory that it was perceived the connection between the appearance of divergent quantities and the product of distributions, whereas distribution theory is strictly linear, as we shall see. This obstacle can be overcome using the idea of extension of distributions, which we will encounter in Sec. \ref{Sec3}. What we shall present here is a direct consequence of the work of Epstein and Glaser \cite{epstein_glaser_1973} and texts which adapted their ideas \cite{gracia_bondia_2002,gracia_lazzarini,grange_uv_2006,Grange2007QuantumFA}. This renormalization procedure is equivalent to the extension of distributions and this is exactly our approach in this paper. Curiously enough, just after the CPT development there were only but a couple of applications, probably due to the necessity of the rigorous functional analysis so applied \cite{prange_epstein-glaser_1999}. Currently, it is more than established. Throughout the method, the renormalization is established by subtracting pertinent test functions, while the physicist community is used to working with distributions in an integral kernel representation. Our work shows (i) that not only does distributions present themselves as good candidates for describing fields, but it also (ii) discusses how a particular scheme of renormalization can be accomplished.

There are different methods for literally ``sweeping the infinities under the rug'' \cite{griffiths_introduction_2011}. In the first step, one \textit{regularizes} the divergent integral, rendering it finite, suppressing the UV physics by a regulator. Then, one posits renormalization conditions, imposing finite conditions on physical quantities. In turn, this defines the cutoff dependence of the bare parameters. After evaluating the now solvable integral, the cutoff parameter is sent to infinity. The result depends of a finite term independent of the regulator plus another one that blows up as the parameter increases. This result is confusing but not that tragic, after all we can only measure the physical values, allegedly finite. The process of the CPT renormalization is managed to take place only once in every step. The lower-order contributions are already renormalized. This corresponds to finding all divergent subgraphs in the traditional approach and simplifies the proof of the same construction to all orders \cite{prange_epstein-glaser_1999}.

In view of this, this paper expresses how distributions are key elements even within the classical regime of electrodynamics. We shall also present, in Sec. \ref{Sec3} a method for extending distributions. Finally, we apply such method, in Sec. \ref{Sec4}, to the renormalization of the self-energy of the electron, which is the main objective of our paper. We point out that, although distributions do not have direct physical meaning, the values that we find upon their application on smooth functions are clearly physical and finite. In particular, the value to the electron self-energy would not have been achieved without the use of distributions (and their extensions).
All this chain was written to be an instructive introduction to the theme. The basic notions about distribution theory, though, are assumed to be known and the interested reader may view a deeper discussion in the existing literature \cite{schwartz_1957, schwartz_mathematics_2008, hormander_analysis_2003, richards_theory_1990, lemos_convite_2013}.

\vskip2ex

\section{Preliminary definitions and results} \label{Sec3}

As we have hinted in Sec. \ref{Sec1}, the developments of QFT showed that renormalization in causal perturbation theory depends heavily on distribution theory, more specifically on the procedures for obtaining extensions of distributions whose behavior at the origin (such nomenclature will become clear later) does not allow that we apply these over functions whose support contains the origin. Our main problem becomes:

\textit{Given a distribution} $T_0$ \textit{which is only well defined when we apply it on test functions} $\varphi \in \eft$ \textit{for which} $0 \notin \textit{supp } \varphi$, \textit{how can we construct a new distribution} $T \in \eftdual$ \textit{which is the extension of} $T_0$, \textit{that is, such that} $\inp*{T_0}{\varphi} = \inp*{T}{\varphi}$ \textit{whenever} $0 \notin \textit{supp } \varphi$?

A presentation of the method for constructing such extensions is the goal of this section. The ideas here presented originate mainly from \cite{prange_epstein-glaser_1999,gracia_lazzarini,gracia_bondia_2002} and references therein.

\subsection{Extensions of distributions} \label{Sec3.2}

Consider a regular distribution $T_f = f \in \eft$ such that $f(x) = 0$ for every $x$ belonging to a subset $U \subset \mathds{R}^n$. It is then evident that, for all $\varphi \in \eft$ such that $\textit{supp } \varphi \subset U$ (we write, in this case, $\varphi \in \eftsub{U}$), we have $\inp*{f}{\varphi} = f(\varphi) = 0$. Note that this can be a form of characterizing the support $\textit{supp } f$ of the function $f$. With the intent of extending this notion to distributions $T \in \eftdual$, we say that $T$ is zero in a subset $U \subset \mathds{R}^n$ when
\[ \inp*{T}{\varphi} = 0 \; , \quad \forall \; \varphi \in \eftsub{U}. \]

We are thus able to define the \textbf{support of a distribution $T$}.

\begin{definition}
    For a given $T \in \eftdual$, the support of $T$ is the subset $\textit{supp } {T} \subset \mathds{R}^n$ given by
    \[ \textit{supp } {T} := \{ x \in \mathds{R}^n \; ; \; x \mbox{ does not contain a neighborhood in which $T$ is zero} \}. \]
    In an equivalent manner, $\textit{supp } {T}$ can be seen as the complement of the biggest subset in which $T$ is zero.
\end{definition}

This definition, in turn, allows one to define an important subspace of $\eftdual$. For any subset $U \subset \mathds{R}^n$, the subspace of $\eftdual$ given by the distributions such that $\textit{supp } {T} \subset U$ will be denoted by $\eftdualsub{U}$.

One important example can be given by the Dirac delta distribution, for which we have $\textit{supp } {\delta_{x_0}} = \{ x_0 \}$. We can actually obtain a reciprocate from this result with the following lemma (for the proof, see for example \cite{hormander_analysis_2003}).

\begin{lemma} \label{le10}
    If $T \in \eftdual$ is such that $\textit{supp } T = \{x_0\}$, then there exists $m \in \mathds{N}$ and constants $c_{\nu}$, $\norma{\nu} \leq m$, such that
    \begin{equation}
        T = \sum_{\norma{\nu} \leq m} c_\nu D^\nu \delta_{x_0}.
    \end{equation}
\end{lemma}

In what follows, we define a quantity which probes the behavior of a distribution $T$ in the origin, in terms of singularities. Since we have drawn a clear distinction between \textit{bona fide} functions (regular distributions) and general (singular) distributions, we must develop for the latter the notion of studying the behavior of $T$ at some point in $\mathds{R}^n$. It is in this ground that we introduce the concept of a \textit{pull-back}.

\begin{definition} \label{def2}
    The \textbf{pull-back of $T \in \eftdual$} over an invertible\footnote{The use of $\Phi$ invertible is a particular case of a definition that can be made more general. More details can be found in \cite{hormander_analysis_2003}.} transformation $\Phi : \mathds{R}^n \rightarrow \mathds{R}^n$ is a new distribution $\Phi^* T$, defined by\footnote{Here the symbol $\norma{D F (y)}$ represents the Jacobian of $F : \mathds{R}^n \longrightarrow \mathds{R}^n$.}
    \begin{equation} \label{eq15}
        \inp*{\Phi^* T}{\varphi} = \inp*{T}{\norma{D \Theta(y)} \Theta^* \varphi} \; , \forall \; \varphi \in \eft,
    \end{equation}
    where $\Theta = \Phi^{-1}$. It is common the notation $T(\Phi(x))$ (which we shall employ from here on, for conformity) for the distribution $\Phi^* T$, making reference to the definition of the pull-back of a function.
\end{definition}

\begin{obs}
    Let us take here some more lines to achieve a better understanding of Definition \ref{def2}. Despite the indication given by the notation $T(\Phi(x))$, the pull-back of a distribution should not be read as an ordinary function. If $T$ is singular, then $\Phi^* T$ is only well defined as a new distribution, which will again be singular. Thus, we should not interpret $T(\Phi(x))$ as an object which varies with $x \in \mathds{R}^n$, but instead as a distribution dependent on the transformation $\Phi$. Nonetheless, when regarding simple (but important) cases, such as $\Phi(x) = \lambda x$, it is easier and perhaps more didactic to express the transformation directly as the argument of $T$.
    
\end{obs}

After such remarks, we present the following definition.

\begin{definition}
    Let $T \in \eftdual$ be a distribution. The \textbf{scaling degree of $T$} is the real number (or $\pm \infty$), denoted here by $\sigma(T)$, such that
    \[ \sigma(T) = \inf \{ s \in \mathds{R} \; ; \; \lambda^{s} T(\lambda x) \xrightarrow{\lambda \to 0^+} 0 \}.\footnote{The convergence to be read in the sense of distributions.} \]
    The \textbf{singular order of $T$}, denoted by $\omega (T)$, is the value
    \[ \omega (T) = [\sigma (T)] - n,\]
    where $[m]$ denotes the biggest integer smaller (or equal) to $m$.
\end{definition}

\begin{example} \label{ex12}
    It is instructive to pass through some examples of distributions and their respective scaling degrees, in the hopes of making our definitions clearer. The examples will be useful to our further discussions as well. 
    
    \begin{enumerate}
        \item If $T = \delta \in \mathcal{D}' (\mathds{R})$, then $\sigma(\delta) = 1$. This is given by the result $\delta(\lambda x) = \lambda^{-1} \delta(x)$, which follows directly from \eqref{eq15}. More generally, dealing with the $n$-dimensional case, $\delta \in \eftdual$, we have
        \[ \norma{D \Phi^{-1}(y)} = \lambda^{-n}, \mbox{ if } \Phi(x) = \lambda x, \]
        implying that
        \[ \inp*{\delta(\lambda x)}{\varphi} = \lambda^{-n} \inp*{\delta}{\varphi}. \]
        This means that $\sigma(\delta) = n$.
        
        \item If $f$ is a continuous function, of one variable, homogeneous with degree $m$, meaning $f(\lambda x) = \lambda^m f(x)$, then
        \[\inp*{\lambda^s f(\lambda x)}{\varphi} = \lambda^s \int_{-\infty}^{+\infty} \lambda^m f(x) \varphi(x) dx = \lambda^{s+m} \inp*{f}{\varphi}. \]
        
        It follows immediately from this that $\sigma(f) = -m$. Thus, \textit{the scaling degree of a homogeneous function is the (additive) inverse of its homogeneity degree}.
    \end{enumerate}
\end{example}
\vskip2ex

We now posses the appropriate tools for enunciating the results concerning proper extensions of distributions. We must separate our problem into two cases, which are characterized by $\omega (T) < 0$ and $\omega (T) \geq 0$, that is, the order relation of the scaling degree and the dimension of our space. The biggest difference between the two situations is in the uniqueness of the extension.

\begin{theorem} \label{theo4}
    Let $T_0 \in \eftdualsub{\mathds{R}^n \backslash \{0\}}$ such that $\sigma (T_0) = s < n$. Then there exists a unique distribution $T \in \eftdual$ such that $\sigma (T) = s$ and
    \[ \inp*{T}{\varphi} = \inp*{T_0}{\varphi} \; , \quad \forall \; \varphi \in \eftsub{\mathds{R}^n \backslash \{0\}}. \]
\end{theorem}

Before passing to the next (and arguably more important) result, we must define some more concepts important to our study. We will denote by $\eftomega$ the subspace of $\eft$ composed of functions such that their derivatives up to order $\omega$ vanish in the origin. Thus, for some natural number $\omega > 0$,
\[ \eftomega = \{ \varphi \in \eft \; ; D^{\alpha} \varphi (0) = 0 \; , \norma{\alpha} \leq \omega\}. \]

Now, for every function $\varphi \in \eft$, its Taylor expansion will be given by
\begin{equation}
    \varphi(x) = \sum_{\nu} \frac{x^\nu}{\nu!} D^\nu \varphi(0),
\end{equation}
from which, separating the terms whose multi-index have norm $\norma{\nu} > \omega$, we obtain
\begin{equation} \label{eq22}
    \varphi(x) = \sum_{\norma{\nu} \leq \omega} \frac{x^\nu}{\nu!} D^\nu \varphi(0) + \sum_{\norma{\nu} > \omega} \frac{x^\nu}{\nu!} D^\nu \varphi(0).
\end{equation}

Therefore, the inclusion $\varphi \in \eftomega$ is equivalent to saying that the first summation in \eqref{eq22} is zero.

The projection of $\eft$ onto $\eftomega$ also plays a crucial role to our second result. Let $\mathcal{W} = \eftomega^\bot$ be the orthogonal complement of $\eftomega$ so that $\eft = \eftomega \oplus \mathcal{W}$.\footnote{$\eftomega$ is evidently closed due to the continuity of any $D^\alpha \varphi$.} 
This allows us to write functionals in $\eftdual$ as
\[ T = T_\omega \oplus l \; , \quad T_\omega \in \eftdualomega \; , \quad l \in \mathcal{W}'. \]

We should ask ourselves then what are the functionals in $\mathcal{W}'$. They are characterized as elements $l \in \eftdual$ such that $\inp*{l}{\varphi} = 0$ for any $\varphi \in \eftomega$. More specifically, we have the following.

\begin{lemma} \label{le8}
For an arbitrary real number $\omega > 0$, we have
\[ \mathcal{W}' = \left\{ l \in \eftdual \; ; \; l = \sum_{\norma{\alpha} \leq \omega} c_\alpha D^\alpha \delta \; , \quad c_\alpha \in \mathds{C} \right\}. \]
\end{lemma}

Once $\mathcal{W}'$ is a linear space, it is a consequence of Lemma \ref{le8} that $\mathcal{B} = \{ D^\alpha \delta \; ; \norma{\alpha} \leq \omega \}$ is a basis to such dual space. We use, however, a different set of distributions to represent the orthogonal projection of $\eftdual$ on $\eftdualomega$, one that depends upon a particular test function $w \in \eft$. We can show that, for any function $w \in \eft$ such that $w(0) \neq 0$, the set\footnote{This means the action of an element of $\mathcal{C}$ over a $\varphi \in \eft$ is $\inp*{ D^\alpha \delta (w^{-1} \cdot)}{\varphi} = \inp*{ D^\alpha \delta }{w^{-1} \varphi}$.} $\mathcal{C} = \{ D^\alpha \delta (w^{-1} \cdot) \; ; \norma{\alpha} \leq \omega \}$ is a basis to $\mathcal{W}'$ as well.

The proof of this argument proceeds as follows: the number of elements in $\mathcal{C}$ equals the number of elements in $\mathcal{B}$, so that one need only to show that the elements of the latter may be generated by $\mathcal{C}$. For this we use the Leibniz rule, which implies that, for any multi-index $\alpha$ and any $\psi \in \eft$ that does not vanish in a given point $x \in \mathds{R}^n$,
\[ D^\alpha \varphi (x) = \frac{1}{\psi(x)} D^\alpha (\psi \varphi) (x) - \frac{1}{\psi(x)} \sum_{0 \leq \beta < \alpha} \frac{\alpha!}{\beta ! (\alpha - \beta)!} (D^\beta \varphi) (x) (D^{\alpha - \beta} \psi) (x). \]
Hence, if $\inp*{D^\beta \delta}{\varphi} = (-1)^{\norma{\beta}} D^\beta \varphi (0)$ may be written as a linear combination of terms such as $\inp*{D^\gamma \delta (w^{-1} \cdot)}{\varphi} = (-1)^{\norma{\gamma}} D^\gamma (w^{-1} \varphi) (0)$ for arbitrary\footnote{This inequality means that $\beta_i \leq \alpha_i$ for all $i \in \{1 , \cdots , n\}$ and at least one of its coordinates, $\beta_j$, is strictly lesser than $\alpha_i$.} $\beta < \alpha$, then the same holds for $D^\alpha \varphi(0)$ by an argument of induction. We conclude our proof noting that the result is valid for $\alpha = (0, \cdots, 0)$,
\[ \inp*{D^\alpha \delta}{\varphi} = \varphi (0) = w(0) \frac{\varphi(0)}{w(0)} = w(0) \inp*{D^\alpha \delta (w^{-1} \cdot)}{\varphi}. \]

Just as any element in $\mathcal{B}$ may be written as a linear combination of elements in $\mathcal{C}$, the same is true for $\mathcal{W}'$. In addition, the set $\mathcal{E} = \{ \frac{(-1)^{\norma{\alpha}}}{\alpha!} w(x) x^\alpha \; ; \norma{\alpha} \leq \omega \}$ generates
$\mathcal{W}$ and will be a basis to the dual of $\mathcal{C}$, since\footnote{The Kroenecker delta used here is a generalization for multiple variables, which is $1$ whenever $\alpha = \beta$, and zero otherwise.}
\[ \inp*{D^\alpha \delta (w^{-1} \cdot)}{\frac{(-1)^{\norma{\beta}}}{\beta!} w(x) x^\beta} = \frac{(-1)^{\norma{\alpha} + \norma{\beta}}}{\beta!} (D^\alpha x^\beta)(0) = \delta_{\alpha,\beta}. \]
\vskip1em

Then, we can write the projection operator of $\eft$ onto $\eftomega$ in the form
\begin{align}
        W_{(\omega;w)} \colon \eft & \longrightarrow \eftomega \nonumber \\
        \varphi(x) & \longmapsto \varphi(x) - w(x) \sum_{\norma{\alpha} \leq \omega} \frac{x^\alpha}{\alpha !} \left( D^\alpha \frac{\varphi}{w} \right)(0), \label{eq20}
\end{align}
for any $w \in \eft$ such that $w(0) \neq 0$.

We chose $w^{-1}$ instead of the function $w$ itself because the former allows us to write an interesting and useful property, to be used ahead. Namely, the operator $W$ satisfies 
\begin{equation} \label{eq24}
    W_{(\omega ; w)} (w \varphi) = w W_{(\omega ; 1)} (\varphi).
\end{equation}
We point out that there is no problem at all when using
$W_{(\omega;\psi)}$, with $\psi$ being the constant function equal to unity. Although $W_{(\omega;1)} \varphi$ is not a test function (its support is not compact), it is infinitely differentiable and, hence, its product with  $w$ will be, in fact, an element of $\eft$.

In accordance with \eqref{eq24}, we have, if $\norma{\alpha} \leq \omega$,
\[ W_{(\omega ; w)} (w x^\alpha) = w W_{(\omega ; 1)} x^\alpha, \]
that is,
\[ W_{(\omega ; w)} (w x^\alpha) (x) = w(x) \left[ x^\alpha - \sum_{\norma{\beta} \leq \omega} \frac{x^\beta}{\beta !} \left( D^\beta x^\alpha \right)(0) \right]. \]
The last sum reduces to $x^\alpha$, so that  
\begin{equation} \label{eq23}
    W_{(\omega ; w)} (w x^\alpha) \equiv 0.
\end{equation}

Our next result is the following 
\begin{theorem} \label{theo5}
    Let $T_0 \in \eftdualsub{\mathds{R}^n \backslash \{0\}}$ such that $\sigma (T_0) = s \geq n$ and $\omega = \omega (T_0) = s - n$. Moreover, given $w \in \eft$, with $w(0) \neq 0$, and constants $C^\alpha \in \mathds{C}$ for all multi-index $\alpha$, with $\norma{\alpha} \leq \omega$. There exists one, and only one, distribution $T \in \eftdual$ such that $\sigma (T) = s$ and satisfying
    \begin{enumerate}
        \item $\inp*{T}{\varphi} = \inp*{T_0}{\varphi} \; , \quad \forall \; \varphi \in \eftsub{\mathds{R}^n \backslash \{0\}}$.
        
        \item $\inp*{T}{w x^\alpha} = C^\alpha$.
    \end{enumerate}
    
    Specifically, $T$ is given by
    \begin{equation}
        \inp*{T}{\varphi} = \inp*{T_\omega}{W_{(\omega;w)} \varphi} + \sum_{\norma{\alpha} \leq \omega} \frac{C^\alpha}{\alpha !} \left( D^\alpha \frac{\varphi}{w} \right)(0),
    \end{equation}
    where $T_\omega$ is the only extension guaranteed by the Theorem \ref{theo4} and $W_{(\omega;w)}$ is the operator $W$, defined in \eqref{eq20}.
\end{theorem}

\begin{obs}
    We can simplify even further our calculations of the extension of $T_0$ if we restrict the class of functions permitted to $w$. More specifically, if we take $w$ such that $(D^\alpha w)(0) = \delta_{\alpha,0}$, which is equivalent to taking $w$ equal to 1 in a neighborhood of the origin, we have
    \begin{equation}
        \begin{split}
            (D^\alpha \frac{\varphi}{w}) (0) & = \sum_{\substack{0 \leq \beta \leq \alpha \\ \beta \neq \alpha}} \frac{\alpha!}{\beta ! (\alpha - \beta)!} (D^\beta \varphi) (0) (D^{\alpha - \beta} w^{-1}) (0) \\
            & = D^{\alpha} \varphi (0),
        \end{split}
    \end{equation}
    since any derivation of $w^{-1}$ will also carry some derivative of $w$ itself. From that, it follows that $W$ reduces to
    \[ (W_{(\omega;w)} \varphi)(x) = \varphi(x) - w(x) \sum_{\norma{\alpha} \leq \omega} \frac{x^\alpha}{\alpha !} D^\alpha \varphi (0) \]
    and the extension $T$ given by Theorem \ref{theo5} can, therefore, be written as
    \begin{equation}
        \inp*{T}{\varphi} = \inp*{T_\omega}{W_{(\omega;w)} \varphi} + \sum_{\norma{\alpha} \leq \omega} \frac{C^\alpha}{\alpha !} \left( D^\alpha \varphi \right)(0).
    \end{equation}
    
    Such functions, satisfying $(D^\alpha w)(0) = \delta_{\alpha,0}$, are called Epstein-Glaser functions (see, for example, \cite{grange_uv_2006}).
\end{obs}

\subsection{Dependence of the extension on the test function $w(x)$} \label{Sec3.3}

We have seen that, for the case when $\sigma (T_0) \geq n$, it does not seem possible to get rid of the dependence of the extension $T \in \eftdual$ on the test function $w \in \eft$ chosen to construct the projection of $\eft$ over $\eftomega$. We can, nonetheless, observe the behavior of this dependence, mainly by studying the term $\inp*{T_\omega}{W_{(\omega;w)} \varphi}$, which we denote, following \cite{prange_epstein-glaser_1999}, the \textit{integral kernel of the extension}. Here, we will stick with the supposition that $w$ is an Epstein-Glaser function, in the sense defined above.

Thus, choosing two Epstein-Glaser test functions $w_1,w_2 \in \eft$, we have
\begin{equation}
    \begin{split}
        (W_{(\omega;w_1)} \varphi) (x) & = \varphi(x) - w_1(x) \sum_{\norma{\alpha} \leq \omega} \frac{x^\alpha}{\alpha !} D^\alpha \varphi (0) \\
        & = \varphi(x) - w_2(x) \sum_{\norma{\alpha} \leq \omega} \frac{x^\alpha}{\alpha !} D^\alpha \varphi (0) + (w_2(x) - w_1(x)) \sum_{\norma{\alpha} \leq \omega} \frac{x^\alpha}{\alpha !} D^\alpha \varphi (0) \\
        & = (W_{(\omega;w_2)} \varphi) (x) + (w_2(x) - w_1(x)) \sum_{\norma{\alpha} \leq \omega} \frac{x^\alpha}{\alpha !} D^\alpha \varphi (0)
    \end{split}
\end{equation}
and, applying $T_\omega$ over this expression, we obtain (bearing in mind that $T_\omega$ is unique, by Theorem \ref{theo4})
\begin{equation} \label{eq26}
    \inp*{T_\omega}{W_{(\omega;w_1)} \varphi} = \inp*{T_\omega}{W_{(\omega;w_2)} \varphi} + \sum_{\norma{\alpha} \leq \omega} \inp*{T_\omega}{(w_2(x) - w_1(x))\frac{x^\alpha}{\alpha !}} D^\alpha \varphi (0).
\end{equation}

We therefore conclude that the application of $T_\omega$ over different projections differ only by a linear combination of terms of the form $\inp*{D^\alpha \delta}{\varphi}$, that is, by application, over the test function $\varphi$, of operators belonging to $\mathcal{W}'$.

Our goal ahead will be, however, to get rid of the restriction that $w$ be a test function. As we have already mentioned, this seems to be a crucial condition for defining $W$ as a projection operator, since it is responsible for the fact that the term $w(x) \sum_{\norma{\alpha} \leq \omega} \frac{x^\alpha}{\alpha !} D^\alpha \varphi (0)$ has compact support. However, very often (as will be the case in a while) a distribution which is not well behaved at the origin will behave nicely at infinity. To be more precise, we mean that such distributions will be well defined when applied to functions $\varphi \in C^\infty (\mathds{R}^n)$ whose support may not be compact. In that sense, taking a sequence $(w_k) \subset \eft$ whose pointwise limit $w$ is a function\footnote{Some works, such as \cite{prange_epstein-glaser_1999}, go even further as to only ask that $w \in \eftdual$ be a distribution. We will not need this generality, so that we have preferred to omit this possibility.} in $C^\infty (\mathds{R}^n)$ and $T_0$ is such that
\[ \lim_{k \to \infty} \inp*{T_\omega}{W_{(\omega;w_k)} \varphi} \in \mathds{C}, \quad \forall \quad \varphi \in \eft, \]
then there is no motive to not consider $w$ in our renormalization scheme. We shall see how this is an important part for the application of this method of extending distributions in the following section.

\section{Application to the electron self-energy} \label{Sec4}

We can now finally attack the electron self-energy problem, seen as a point particle. We now posses sufficient machinery to see this task as a pathology to be faced with extension of distributions defined over $\eftsub{\mathds{R}^n \backslash \{0\}}$. To clarify our aims, we will translate the issues of electrostatics to the language of distributions.  

The first and default  example  is the charge  distribution of an electron, seen as a charged point particle, which is represented by the Dirac delta $\delta(\cdot)$, centered where we suppose the whole electric charge of the particle should be concentrated. Actually, this particular case is not the only one where we consider the charge distribution $\rho$ as a generalized function. After all, it would be an incredible coincidence in nomenclature if the charge distributions presented in the electrodynamics were not presented by distributions $\rho \in \efttdual$. This is one of the main contributions of the theory of generalized functions to electrodynamics. Not only does it incorporate more general charge distributions $\rho$ (which could not be defined just as real functions in $\mathds{R}^3$), but it also eases some of its manipulations.
 
\begin{example}
    To illustrate our last paragraph, let us analyze the representation of an electric dipole as a distribution in $\eftt$. The pure dipole, considered an idealization such as point charges, is constructed as follows: start by setting two charges, $q$ and $-q$, separated by a fixed distance $\varepsilon$. Suppose that they lie in the $x$ coordinate, with $-q$ at the origin and $q$ at $x = \varepsilon$, accordingly. That way, the distribution  $\rho \in \eftsub{\mathds{R}}$ will be
    \[ \rho = - q \delta_0 + q \delta_\varepsilon = q (\delta_\varepsilon - \delta_0). \]
    
    If we set $q = 1/\varepsilon$, then, in the limit $\varepsilon \to 0^+$ we obtain a distribution whose total charge vanishes. This result is not correct nonetheless. A null charge distribution would lead to a zero electric field, which is not the case for a dipole (see \cite{griffiths_electro_1999} for the expression of \textbf{$\vec{E}$} in this case). 
    
    This simple argument indicates that the distribution formalism is more than necessary to an accurate description of even the basics of electrostatics. 
    
    Now, we apply $\rho_\varepsilon = \varepsilon^{-1} (\delta_{\varepsilon} - \delta_0)$ over any $\varphi \in \eftsub{\mathds{R}}$. We have
    \[ \inp*{\rho_\varepsilon}{\varphi} = \frac{1}{\varepsilon} \inp*{\delta_{\varepsilon} - \delta_0}{\varphi} = \frac{\varphi(\varepsilon) - \varphi(0)}{\varepsilon} \]
    and, therefore, we shall obtain, in the limit $\varepsilon \to 0^+$,
    \[ \inp*{\rho_\varepsilon}{\varphi} \to \varphi'(0) = - \inp*{\delta'}{\varphi}, \]
    that is,
    \[ \rho_{dip} = - \delta'. \]
    
    The generalization to the three-dimensional case is straightforward. In this case, the dipole moment $\textbf{p} \coloneqq q \vec{\varepsilon}$, where $\vec{\varepsilon}$ is the displacement vector that connects the negative to the positive charge. Once again we take the limit $\vert \vec{\varepsilon} \vert \to 0$. The way we have taken $q$ above, we guaranteed that the dipole moment was kept constant, even when the charges are close enough. 
    With this premise, the charge distribution will be given by
    \[ \inp*{\rho_\varepsilon}{\varphi} \to - \frac{\partial}{\partial \textbf{p}} \varphi(0)\; , \qquad \forall \; \varphi \in \eftt, \]
    that we denote by
    \[ \rho_{dip} = - \textbf{p} \cdot \nabla \delta \in \eftt. \]
\end{example}
\vskip2ex

We now return to the main problem  we intend to treat, namely, the divergence of the electron self-energy. In this context, the fact that the charge is fully concentrated in the origin is seen by the application of $q \delta$ on different test functions $\varphi \in \eftt$. For any test function whose support does not contain the origin, we have   
\begin{equation}
    \int_{\textit{supp } \varphi} \rho(x) \varphi(x) d^3 x = 0.
\end{equation}

At the same time, the distribution possesses a finite charge once the integration of $\rho$ over $\mathds{R}^3$ is equivalent to applying $q \delta$ on a test function $\varphi \in \eftt$ whose support contains the origin such that $\varphi(0) = 1$. 
\begin{equation} \label{eq4.1}
    \int_{\mathds{R}^3} \rho (x) \varphi(x) \; d^3 x = \int_{\textit{supp } \varphi} \rho (x) \varphi(x) \; d^3 x = \inp*{q \delta}{\varphi} = q.
\end{equation}

This charge distribution produces both a potential $V$ and an electric field  $\textbf{E}$, which are also new distributions. We haven't considered vector fields as distributions so far, just like $\textbf{E}$, but this generalization is quite natural and $\textbf{E}$ acts on an element of $\eftt$ according to  
\[ \inp*{\textbf{E}}{\varphi} = (\inp*{E_x}{\varphi}, \inp*{E_y}{\varphi}, \inp*{E_z}{\varphi}). \]

The application of vector fields as distributions were mentioned here just for completeness. It will be no longer necessary hence forth. 

We can see that, in fact, $V$ represents an element of $\efttdual$. The explicit formula for the potential is given by $V(\textbf{r}) = \frac{e}{r}$, where $e$ is the electron charge and we are using unities in which $4 \pi \epsilon_0 = 1$, with no further implications to the final results whatsoever. As usual, $r$ represents the radial coordinate of a spherical coordinate system centered on the charge. $V(\cdot)$ is a smooth function for any $\textbf{r} \neq 0$. Hence, we just have to be concerned  to the convergence of $\inp*{V}{\varphi}$, for an arbitrary $\varphi \in \eftt$. In effect, if $R > 0$ is such that $K = B_R(0) \supset \textit{supp }\varphi$  and $M = \max_{x \in \mathds{R}^3}{\varphi(x)}$, then
\begin{equation}
    \begin{split}
        \norma{\int_{\mathds{R}^3} V(x) \varphi (x) d^3 x} & \leq M \int_{K} V(x) d^3 x \\
        & = M \left( \int_0^{2 \pi} d \phi \int_0^\pi \sin{\theta} d \theta \right) \int_{0}^R \frac{e}{r} r^2 dr \\
        & = (2 \pi M e) R^2 < \infty.
    \end{split}
\end{equation}

An analogous consideration may be done for the electric field, since  $\textbf{E} \sim \frac{1}{r^2}$. In view of that, the radial integral will converge as well. We can thus turn our attention to the stored self-energy of a charged system. As we have already seen, for the particular case of an electron there is a divergence at the origin. We are considering here only the static case, so we don't have to worry about  magnetic fields. We can consider, however, other cases where such divergence does not appear and we are thus able to calculate the system self-energy. If we consider, for example, the electron as a uniformly charged spherical shell of radius $a$, then  
\begin{equation}
    \textbf{E}(\textbf{r}) =
    \begin{cases}
          0 & \mbox{, se } r \leq a, \\ 
          (e/r^2) \; \hat{\textbf{r}} & \mbox{, se } r > a. \\
    \end{cases}
\end{equation}
Therefore, the self-energy, see eq. \eqref{eq1.6}, will be given by
\begin{equation}
W = \frac{1}{8 \pi} \int_{\mathds{R}^3} \textbf{E}^2 d \tau = \frac{1}{8 \pi} (4 \pi) \int_a^\infty \frac{e^2}{r^2} d r = \frac{1}{2} \frac{e^2}{a}.    
\end{equation}

In distribution parlance, the last equation is just the application of the regular distribution $\frac{1}{8 \pi} \textbf{E}^2$ over the function $\varphi(x) \equiv 1$. We point out that, even with $\varphi \notin \eftt$, $\inp*{\textbf{E}^2}{\varphi}$ does exist. It is a consequence of the behavior of $\textbf{E}^2$ at infinity, which is good enough that we do not need to restrict the range of integration to a compact set. For general distributions, with unknown behavior at infinity, this restriction is considered by supposing that $\varphi$ is a test function.   

We have already mentioned, see Sec. \ref{Sec3.3}, that it is often advantageous (or even necessary) to work with functions that are not compactly supported. This is allowed once our distribution possesses the necessary conditions so that its application over this larger class of functions is well behaved. For instance, the Dirac delta may be applied on any function $\varphi$ continuous at the origin. In our specific case, we will see how the electron self-energy ($\sim \textbf{E}^2$) behavior far from the origin permits such loosening of the conditions over $\varphi$. More precisely,  
\begin{equation} \label{eq27}
    \textbf{E}^2 = \frac{e^2}{r^4}, \quad r > 0
\end{equation}
is not well defined as a distribution. In effect, for any test function obeying $\varphi(x) = 1$ for $x$ in a neighborhood $\mathcal{V}$ of the origin, say, a ball, we have  
\[ \inp*{\textbf{E}^2}{\varphi} = \int_{\mathcal{V}} \frac{e^2}{r^4} d^3 x + \int_{\mathds{R}^3 \backslash \mathcal{V}} \frac{e^2}{r^4} \varphi(x) d^3 x = + \infty, \]
once the first term is linear divergent due to the fourth-order homogeneity of $\textbf{E}^2$.\footnote{The classification as a linear divergence may be justified in polar coordinates. Taking $R = 1/r$, we find $\int_0^{+\infty} \frac{dr}{r^2} = \int_0^{+\infty} dR$. See \cite{Bovy:2006dr} for details.}   

Outside the origin, however, $\textbf{E}^2$ is a smooth function, and as such, locally integrable. Hence, we have, at least, $\textbf{E}^2 \in \eftdualsub{\mathds{R}^3 \backslash \{ 0 \}}$. For this reason, we may extend (renormalize) the distribution  $\textbf{E}^2$ to a new distribution $U \in \eftdual$ with the methods exposed previously in Sec. \ref{Sec3}. In that way, we expect that the electron self-energy $E_0$ will be well defined as the application of $\frac{1}{8\pi} U$ over the function $\varphi \equiv 1$, 
\begin{equation} \label{eq31}
    E_0 = \frac{1}{8\pi} \inp*{U}{1}.
\end{equation}

According to what we have made so far, let us first determine the scaling degree and the singular order of $\textbf{E}^2$, which are key to Theorems \ref{theo4} and \ref{theo5}. We observe that it is a homogeneous function of order $-4$, so that, as expressed through the Example \ref{ex12},
\[ \lambda^s \textbf{E}^2 (\lambda x) = \lambda^s \frac{e^2}{(\lambda r)^4} = \lambda^{s-4} \textbf{E}^2, \]
which implies,
\[ \lambda^s \textbf{E}^2 (\lambda x) \xrightarrow{\lambda \to 0^+} 0 \iff s > 4. \]
That is, $\sigma (\textbf{E}^2) = 4$, and also 
$\omega (\textbf{E}^2) = \sigma (\textbf{E}^2) - n = 1$.

Therefore, for a test function $w \in \eftt$ and constants $C^0,C^{(1,0,0)} \equiv C^1$, $C^{(0,1,0)} \equiv C^2$, $C^{(0,0,1)} \equiv C^3$, we obtain, according to the Theorem \ref{theo5}, a distribution
$U \in \efttdual$ defined by 
\begin{equation}
    \begin{split}
        \inp*{U}{\varphi} & = \inp*{\textbf{E}_1^2}{W_{(1;w)} \varphi} + \sum_{\norma{\alpha} \leq 1} \frac{C^\alpha}{\alpha!} \left( D^\alpha \frac{\varphi}{w} \right)(0) \\
        & = \inp*{\textbf{E}_1^2}{W_{(1;w)} \varphi} + C^0 \frac{\varphi(0)}{w(0)} + \sum_{i=1}^3 C^i \partial_{x_i} \left( \frac{\varphi}{w} \right) (0)
    \end{split}
\end{equation}
satisfying
\begin{equation}
    \inp*{U}{\varphi} = \inp*{\textbf{E}^2}{\varphi} \; , \quad \forall \quad \varphi \in \eftsub{\mathds{R}^3 \backslash \{0\}},
\end{equation}
\begin{equation}
    \inp*{U}{w x^\alpha} = C^\alpha \; , \quad \norma{\alpha} \leq 1.
\end{equation}

Moreover, if $w$ is chosen as an Epstein-Glaser function, we have
\begin{equation} \label{eq25}
    \inp*{U}{\varphi} = \inp*{\textbf{E}_1^2}{W_{(1,w)}\varphi} + C^0 \varphi(0) + \sum_{i=3} C^i (\partial_{x_i} \varphi)(0).
\end{equation}

Before moving on, there is a comment in order. The charge distribution of a (charged) point particle possesses spherical symmetry. It is not only due to the concentration of charge in a single point. In fact, a electric dipole has also a distribution whose support is contained in the origin, although the allegedly spherical symmetry is broken once the moment $\textbf{p}$ defines a privileged direction. The additional fact that our point particle model admits no internal structure imposes the constraint of having no special direction.      

Since we would like to preserve such symmetry when extending $\textbf{E}^2$, we must choose $C^i = 0$, for $i = 1,2,3$. This is justified because the last term in \eqref{eq25} does not behave like a scalar under rotations of our coordinate system, unless the three constants vanish. We can promptly see this by writing  
\[ \sum_{i=3} C^i (\partial_{x_i} \varphi)(0) = \textbf{C} \cdot (\nabla \varphi)(0) \; , \quad \textbf{C} = (C^1 , C^2 , C^3). \]
Now, $\nabla \varphi$ does behave as a vector, however we cannot say the same for $\textbf{C}$. For this reason, the only way to keep this sum inert under rotations is to set $\textbf{C} = \textbf{0}$.

From this, we may rewrite $U$ as
\begin{equation}
    \inp*{U}{\varphi} = \inp*{\textbf{E}_1^2}{W_{(1;w)} \varphi} + C^0 \varphi(0).
\end{equation}

Then, we seek, analogous to what has been done in Sec. \ref{Sec3.3}, to relax the conditions imposed on $w$ in the renormalization of $\textbf{E}^2$. Our path will be to take a sequence of test functions $w_M$ that converges pointwise to $w(x) = 1 \in C^\infty (\mathds{R}^3)$, obtaining a well defined distribution given by
\begin{equation} \label{eq30}
    \inp*{U}{\varphi} = \lim_{M \to \infty} \inp*{\textbf{E}_1^2}{W_{(1;w_M)} \varphi} + C^0 \varphi(0).
\end{equation}

The last equation suggests that we will take all $w_M$ as Epstein-Glaser functions. Specifically, each $w_M$ shall be taken as
\begin{equation}
    w_M(x) =
    \begin{cases}
        1, & \mbox{if } x \in B_M(0), \\ 
        \chi(r-M), & \mbox{if } x \in B_{M+1}(0) \backslash B_M(0), \\
        0, & \mbox{if } x \notin B_{M+1}(0),
    \end{cases}
\end{equation}
where $\chi$ is a smooth radial function such that $\norma{\chi(s)} \leq 1$, $s \in [0,1]$, $\chi(0) = 1$ and $\chi(1) = 0$,

Hence, we have $(D^\alpha w_M) (0) = \delta_{\alpha,0}$ and $w_M(x) \rightarrow 1$ for any $x \in \mathds{R}^3$ in the limit $M \to \infty$. 
Moreover, due to the result \eqref{eq26}, for any two naturals $M_1, M_2 \in \mathds{N}$ (say, $M_1 < M_2$), we have
\[ \inp*{\textbf{E}_1^2}{W_{(1;w_{M_2})} \varphi} = \inp*{\textbf{E}_1^2}{W_{(1;w_{M_1})} \varphi} + \sum_{\norma{\alpha} \leq 1} \inp*{\textbf{E}_1^2}{(w_{M_1}(x) - w_{M_2}(x))\frac{x^\alpha}{\alpha !}} D^\alpha \varphi (0), \]
wherein 
\begin{equation}
    w_{M_1}(x) - w_{M_2}(x) =
    \begin{cases}
        0, & \mbox{if } x \in B_{M_1}(0), \\ 
        \chi(r - M_1) - 1, & \mbox{if } x \in B_{M_1 + 1}(0) \backslash B_{M_1}(0), \\
        - 1, & \mbox{if } x \in B_{M_2}(0) \backslash B_{M_1 + 1}(0), \\
        - \chi(r - M_2), & \mbox{if } x \in B_{M_2 + 1}(0) \backslash B_{M_2}(0), \\
        0, & \mbox{if } x \notin B_{M_2+1}(0).
    \end{cases}
\end{equation}

That way, if we denote $(a_M)$ the real sequence whose elements are $\inp*{\textbf{E}_1^2}{W_{(1;w_{M})} \varphi}$, then we will show that it converges, proving that $(a_M)$ is a Cauchy sequence. Indeed, we have just seen that\footnote{Since $w_{M_1}(x) - w_{M_2}(x) \in \eftsub{\mathds{R}^n \backslash \{ 0 \}}$, the action of $\textbf{E}_1^2$ may be replaced by $\textbf{E}^2$.}
\begin{equation}
    \begin{split}
        a_{M_1} - a_{M_2} & = \inp*{\textbf{E}_1^2}{W_{(1;w_{M_2})} \varphi} - \inp*{\textbf{E}_1^2}{W_{(1;w_{M_1})} \varphi} \\
        & = \sum_{\norma{\alpha} \leq 1} \inp*{\textbf{E}^2}{(w_{M_1}(x) - w_{M_2}(x))\frac{x^\alpha}{\alpha !}} D^\alpha \varphi (0),
    \end{split}
\end{equation}
so that, if we limit this sum, then we will also limit the difference $a_{M_1} - a_{M_2}$. Now, given $\varepsilon > 0$, we take $M \in \mathds{N}$ such that $\frac{1}{M} < \frac{\varepsilon}{8 \pi e^2}$ and $M_1, M_2 \in \mathds{N}$, with $M \leq M_1 < M_2$. Since $\textbf{E}^2$ acts on test functions whose support does not contain the origin, we can employ the formula \eqref{eq27}, obtaining
\[ \inp*{\textbf{E}^2}{(w_{M_1}(x) - w_{M_2}(x))} = \int_{\mathds{R}^3} \frac{e^2}{r^4} (w_{M_1}(x) - w_{M_2}(x)) d^3 x, \]
\[ \inp*{\textbf{E}^2}{(w_{M_1}(x) - w_{M_2}(x))x_i} = \int_{\mathds{R}^3} \frac{e^2}{r^4} (w_{M_1}(x) - w_{M_2}(x)) x_i d^3 x. \]

Now, once \textbf{i.} $w_{M_1} - w_{M_2}$ is radial, \textbf{ii.} has support in $B_{M_2 + 1} (0) \backslash B_{M_1} (0)$ and \textbf{iii.} assumes values only in $[0,1]$, we have
\[ \norma{\inp*{\textbf{E}^2}{(w_{M_1}(x) - w_{M_2}(x))}} \leq (4 \pi e^2) \int_{M_1}^{M_2+1} \frac{1}{r^2} dr = (4 \pi e^2) \left[ \frac{1}{r} \right]_{M_2 + 1}^{M_1} \leq (8 \pi e^2) \frac{1}{M_1}, \]
that is,
\begin{equation} \label{eq28}
    \norma{\inp*{\textbf{E}^2}{(w_{M_1}(x) - w_{M_2}(x))}} \leq \frac{8 \pi e^2}{M} < \varepsilon.
\end{equation}

Meanwhile,
\begin{equation} \label{eq29}
    \inp*{\textbf{E}^2}{(w_{M_1}(x) - w_{M_2}(x))x_i} = \int_{\mathds{R}^3} \frac{e^2}{r^4} (w_{M_1}(r) - w_{M_2}(r)) x_i d^3 x = 0,
\end{equation}
because for any $i = 1,2,3$, corresponding to the three Euclidean axis $x,y,z$, respectively, the integration in $\phi$ or in $\theta$ will vanish.\footnote{This is  only possible because $w_M$ is a sequence of radial functions and as such, the integration in  $r, \phi$ and $\theta$ can be factored.} In fact, for $i=1$ and $i=2$, the integrals in $\phi$ vanish, $\int^{2\pi}_0 d\phi \cos \phi =\int^{2\pi}_0 d\phi \sin \phi = 0 $. Now, for $i=3$, we find $\int^\pi_0 d \theta \cos \theta \sin \theta = 0$.

Hence, with the help of \eqref{eq28} and \eqref{eq29}, we may conclude that
\begin{equation}
    \norma{a_{M_1} - a_{M_2}} \leq \varepsilon \varphi(0),
\end{equation}
which, in turn, implies that $(a_M)$ will be a Cauchy sequence, that is, a convergent one.

All the previous development allows us to show  that the distribution defined in \eqref{eq30}, that we will denote simply  by
\begin{equation}
    \inp*{U}{\varphi} = \inp*{\textbf{E}_1^2}{W_{(1;1)} \varphi} + C^0 \varphi(0),
\end{equation}
is well defined. Thus, we may finally use the eq. \eqref{eq31} to obtain the renormalized electron self-energy,
\begin{equation} \label{eq32}
    E_0 = \frac{1}{8 \pi} \inp*{\textbf{E}_1^2}{W_{(1;1)} 1} + \frac{1}{8 \pi} C^0 = \frac{C^0}{8 \pi}.
\end{equation}

Although simple, the eq. \eqref{eq32} bears great physical meaning, concentrating our results. We have shown that defining the electron self-energy as the application of the distribution  $\frac{1}{8 \pi} U \in \eftt$, which extends (or renormalize) $\textbf{E}^2$, over the constant function $1$, we get rid of the divergence previously depicted. This divergence appeared when one directly considers $\textbf{E}^2$, which cannot be seen as an actual distribution.\footnote{The statement that $\textbf{E}^2 \notin \eftt$ is a consequence that the product between distributions is not, in general, well defined. An alternative method to skirt the electron self-energy divergence is related to a generalization to the very concept of distributions, working with the \textit{generalized Coulombeau functions}.  For details, see  \cite{Gsponer_2008,gsponer_concise_2009} and references therein.} Now, $E_0$ becomes an undetermined constant, that we may control to serve our model. This is the very kernel of a renormalization method. If, for instance, we assume that the self-energy is, alone, responsible for the electron mass, we shall take 
\[ C^0 = 8 \pi m_e c^2, \]
in a way that $E_0 = m_e c^2$.

To summarize, we have seen how the self-energy problem originates in the fact that $\textbf{E}^2$ is not a proper element of $\eft$. On the other hand, $\textbf{E}^2$ is, outside the origin, a smooth function. Thus, we at least have $\textbf{E}^2 \in \eftdualsub{\mathds{R}^n \backslash \{ 0 \}}$, which means we can use Theorem \ref{theo5} to extend it to a distribution in $\eftdual$.

\section{Conclusion}
\label{Sec5}

The main objective of this work was to analyze (and renormalize) a simple but central problem in classical electrodynamics: the electron self-energy. Although the electrostatics model of a charged point particle implies a linear divergence on the self-energy, we may skirt this infinity with an extension of the corresponding distribution. With more details,

\textbf{1.} We have shown how distributions and also their extensions can play a central role already in classical electrodynamics. Even the most basic physical quantities such as charge density, scalar and vector fields turn out to be described as linear functionals. In addition, we have also demonstrated that a specific scheme of renormalization, seen as an extension of a particular distribution, can be applied to overcome the divergent electron self-energy.
    
\textbf{2.} The leading results of extension of distributions, that is, the corresponding renormalization, were all enunciated, following the lines in \cite{fredenhagen_notes}. We have focused in a particular subspace of the set of test functions, namely, the one whose elements vanish in an arbitrary neighborhood of the origin. We have investigated the behavior of different distributions in the origin and how one could recover such distributions, in the sense of making them continuous linear functionals defined over all the space of test functions. 
    
\textbf{3.} We have applied the concepts of distributions and the corresponding extensions to the classical electron self-energy. At first sight, electrostatics implies a divergence once we treat the electron as a charged point particle. However, our construction shows that its self-energy turns out to be an undetermined constant upon renormalization, so that our parameters might be fixed, for example, appealing to empirical results.

In conclusion, although the use of distribution-valued fields (and their renormalization as extensions) in classical electrodynamics may seem abstract and disconnected from physical reality, it is an important mathematical tool that enables us to describe the behavior of physical fields in a rigorous and consistent manner. The key is to choose renormalization conditions that correspond to physical observables. In our case, we are using the energy, which could potentially be measured and accepted as an observable.



\end{document}